\documentclass[a4paper,11pt]{article}
\usepackage{jcappub} 
\usepackage{graphicx} 
\usepackage{amsmath}
\usepackage{mathrsfs}
\usepackage{amsfonts}
\usepackage{amsmath}
\usepackage{amssymb}
\usepackage{array}
\usepackage{verbatim}
\usepackage{bm}
\usepackage{epsfig}
\usepackage{graphicx,color}
\usepackage{relsize}
\usepackage{lineno}
\usepackage{float}
\usepackage{multirow}
\usepackage{gensymb}
\RequirePackage{xspace}
\usepackage{graphicx}
\usepackage{amsfonts}
\usepackage{hyperref}
\usepackage{subfigure}
\usepackage{xcolor}
\usepackage{orcidlink}

\usepackage{graphicx}
\usepackage{subcaption}

\newcommand{\plotfromdir}[1]{%
    \centering

    \begin{subfigure}{}
        \includegraphics[width=0.45\linewidth]{#1/subplot_0_0.png}
    \end{subfigure}
    \begin{subfigure}{}
        \includegraphics[width=0.45\linewidth]{#1/subplot_1_0.png}
    \end{subfigure}

    \begin{subfigure}{}
        \includegraphics[width=0.3\linewidth]{#1/Sig_back_x.png}
    \end{subfigure}
    \begin{subfigure}{}
        \includegraphics[width=0.3\linewidth]{#1/Sig_back_y.png}
    \end{subfigure}
    \begin{subfigure}{}
        \includegraphics[width=0.3\linewidth]{#1/Sig_energy.png}
    \end{subfigure}

    \begin{subfigure}{}
        \includegraphics[width=0.3\linewidth]{#1/Obs_x.png}
    \end{subfigure}
    \begin{subfigure}{}
        \includegraphics[width=0.3\linewidth]{#1/Obs_y.png}
    \end{subfigure}
    \begin{subfigure}{}
        \includegraphics[width=0.3\linewidth]{#1/background_energy.png}
    \end{subfigure}

}

\title{Convolutional non-parametric Gamma-Ray Signal and Background Separation} 
\author[a]{Scarlet Betterman}
\author[b]{Emmanuel Moulin}
\author[a]{Martin White}

\affiliation[a]{ARC Centre of Excellence for Dark Matter Particle Physics \& CSSM,
Department of Physics, Adelaide University, Adelaide, SA 5005, Australia}
\affiliation[b]{Irfu, CEA Saclay, Université Paris-Saclay, F-91191 Gif-sur-Yvette, France}

\emailAdd{scarlet.betterman@adelaide.edu.au}
\emailAdd{emmanuel.moulin@cea.fr}
\emailAdd{martin.white@adelaide.edu.au}

\date{\today}

\begin{document}
\abstract{In very-high-energy gamma-ray astronomy, signals must be separated from the residual background which arises from misidentified cosmic-ray protons. Building on previous work, we introduce an ensemble of variational autoencoders that aims to perform automatic signal-background separation with minimal assumptions that include separability of the spatial and energy distributions in both the signal and background, with no prior specification of the number of components or the point-like/diffuse nature of the signal itself. In addition, we do not assume knowledge of which region of the coordinate space is background-dominated or where the signal is supposed to be located in the field of view. We test the model on an analytic point-source mixture scenario, a realistic simulation of dark matter annihilation in the Galactic centre, and real observations of the Crab nebula and MSH 15-52 from the public H.E.S.S data release. The model proves capable of completely reconstructing the signal and background at a pixel by pixel level in all scenarios, whilst also denoising the inputs. Stable performance and reasonable error estimates are obtained even for low signal-to-background ratios.}


\maketitle

\section{Introduction}
The observation of very-high-energy (VHE, E $\gtrsim$ 100 GeV) gamma rays is an important mechanism for studying the most violent phenomena in the universe. Since gamma rays propagate without deflection, they can be traced to their origin to provide insight into high energy events such as the acceleration of relativistic particles within shocks and the annihilation of dark matter. Close analysis of these signals, including a detailed mapping of their spatial, spectral and temporal characteristics, is required in order to characterise these events and their role within the various phenomena of interest. 

For VHE gamma rays, imaging atmospheric Cherenkov telescopes (IACTs), such as H.E.S.S.~\cite{HESS} and the upcoming CTAO~\cite{CTAO}, offer a powerful approach. These telescopes provide highly accurate measurements of gamma rays as they interact with the earth's atmosphere producing Cherenkov air-showers. One unfortunate side effect of IACTs is that observations become contaminated with background cosmic-ray and diffuse gamma-ray signals. Although some amount of cleaning can be done at a per gamma/cosmic-ray level, a certain degree of contamination is unavoidable, especially for observations where the signal of interest is close to, or even below, the level of the background. 

\newpage
To handle this contamination, post-processing on the gamma-ray data is required. Methods for doing this generally rely on a statistical analysis of the overall observation to separate background events from events of interest. A wide variety of methods have been devised for this depending on whether the position of the source is known \emph{a priori} or not. One popular approach is to divide the field of view into an \emph{ON} region (which contains the source signal), and one or more \emph{OFF} regions (which are known to contain only background). For example, the \emph{Ring Background} method defines a ring-shaped region surrounding the \emph{ON}-source area~\cite{Berge:2006ae}, and the \emph{Adaptive ring} method adapts the radius of the ring if it overlaps with known or suspected VHE emission regions~\cite{HESS:2018pbp}. One can also use \emph{multiple} \emph{OFF} regions to obtain more accurate background estimates by averaging over the Poisson fluctuations expected from observations in any given single region~\cite{Berge:2006ae}. Another approach, the \emph{template-background method}, utilises a detailed model of the expected background derived from multiple observations across the field of view~\cite{Rowell:2003jb}. While these methods are all effective at reducing background contamination, each has its drawbacks related to factors such as the computational expense and necessary assumptions about the prior knowledge of the background. 

The problem of separating interesting signals from the cosmic-ray induced background is an ideal use case for machine learning algorithms. Ref.~\cite{Ullmo:2024odq} presented a proof of principle based on training \emph{n} sets of neural networks to estimate the spatial and energy components for two fields corresponding to a typical source and background. Their sum is then constrained via a likelihood maximisation to match the total observation, and subjected to various optional additional constraints such as imposing a threshold on the spatial background intensity, or insisting that the networks reproduce the observed results in a dedicated OFF region. Whilst proving successful on simple test examples, the approach suffered from various limitations arising from the input assumptions. Most severely, the method assumed that there were exactly two components in the data (i.e. one point-like, extended or diffuse signal plus background), leading to trained models that do not generalise well to observations with a different source multiplicity. Furthermore the background threshold constraint presupposes that the background observation is relatively homogeneous, an assumption which is likely to fail in large grid-surveys where inhomogeneities can arise due to varying conditions such as weather, changing azimuthal angles and the varying energy acceptance with respect to the observation zenith angle.

To address these issues, the present work introduces a new convolutional variational autoencoder (VAE) approach to separating background from signal, assuming that the input data consists of binned photon counts in both spatial location and photon energy. Our present work improves over the previous one in the following aspects.
\textit{(i)} The source model is able to generalise to unseen observations as long as the training dataset is sufficiently broad in construction.
\textit{(ii)} No physical prior or knowledge about the observations is assumed, such as what the background-dominated region of the coordinate space is or where the signal is supposed to be located in the field of view. This means the model is completely blind as to what the observation contains in advance.
\textit{(iii)} The approach performs well with low signal-to-background ratios provided that the training dataset is sufficiently built to cover such scenarios.
\textit{(iv)} The number of signal components is left free and the model works for overlapping extended sources in the coordinate space.
\textit{(v)} Our approach is able to de-noise the signal and background components to remove the effect of statistical fluctuations due to Poisson noise and recover mean fields for the different components.

\newpage
The paper is organised as follows. Section~\ref{sec:setup} 
describes the problem setup and outlines our variational autoencoder network architecture. Section~\ref{sec:dataset} presents the data sets on which the model is applied. The results are discussed in 
section~\ref{sec:results}. Section~\ref{sec:discussion} is devoted to  the conclusions of this work with a discussion of the potential outlook.

\section{Setting up the Problem}
\label{sec:setup}
\subsection{Formalism and Method} 
Neural networks~\cite{Schmidhuber:2014bpo} perform efficiently in signal separation in a variety of domains including astronomy~\cite{burke2019deblending,wang2022galaxy}. In Ref.~\cite{Ullmo:2024odq}, 
a neural network-based analysis framework was used to obtain the signal and background contributions to a 2-component mixed astrophysical observation in gamma rays, 
where the signal component was chosen to be either a point source, an extended source, or a diffuse signal such as that arising from dark matter annihilation. In this work, we instead take the following approach: for a 3D data cube of counts binned in the spatial and spectral coordinate space,

we aim to obtain the most probable background and signal distributions for a signal with an arbitrary number of components. This allows a signal to be extracted from observed data with statistical noise removed during the extraction process, without making any assumptions about the type of signal present in the data. It is also extendable to real-time processing of new observations, whilst allowing for fine separation of faint signals mixed with stronger point sources., however exploration of this functionality is left for future work.

Following Ref.~\cite{Ullmo:2024odq}, we assume that an observation can be considered as a 3D array of count events given in the binned $(e,\overrightarrow{x})$ coordinates, where $e$ and $\overrightarrow{x}$ represent energy and 2D space coordinates, respectively. This 3D array is then considered to be a particular Poisson realisation of an underlying mean field $F_{tot}$. Under such a construction, $F_{tot}$ can be assumed to be made of two, or more, components, each the Poisson realisation of its own underlying field. The simplest such decomposition is the separation into a purely signal component ($F_{s}$) and a purely background component ($F_{b}$). It is then further assumed that $F_{s}$ and $F_{b}$ can both be factorised into uncorrelated energy and spatial components such that:
\begin{align*}
    \begin{cases}
       F_{s}(\overrightarrow{x}, e)=S_s(\overrightarrow{x}) \times E_s(e)\, ,\\
       F_{b}(\overrightarrow{x}, e)=S_b(\overrightarrow{x}) \times E_b(e)\, \\
    \end{cases}
\end{align*}
with the total mixture given by:
\begin{equation}
\label{eq:ftot}
    F_{tot}(\overrightarrow{x}, e) = \sum_{i=1}^{n-1} S_{s_i}(\overrightarrow{x}) \times E_{s_i}(e) + S_b(\overrightarrow{x}) \times E_b(e) \, ,
\end{equation}
where the sum runs over the number of signal components $n-1$.

\subsection{Convolutional Neural Network Framework and Architecture}

Our model considers the problem of separating signal from background as two separate reconstruction tasks. Convolutional variational autoencoders provide a strong framework for reconstructing an input in a lossy way that can be manipulated to provide a wide variety of functionality~\cite{DBLP:journals/corr/abs-1906-02691,kingma2013autoencoding}. For example, in image cleaning applications, the autoencoder's training data is modified to reconstruct a clean version of the input. In the 3D histogram representation of our gamma-ray observations, a single observation is not dissimilar to the kind of images found in medical contexts, with the key difference being that the 3rd dimension is spatial rather than spectral. In this way, the 3D array of data is more akin to a coloured image with many extra colour channels. Regardless of this, signal separation using an autoencoder is very analogous to an image cleaning task where the model is trained only to reconstruct the signal from the image. This is not a simple task, since the signal and background are often intertwined in gamma ray images and, in observations over a large field of view, multiple signals may be present in the region of interest. However, another feature of autoencoders makes them a particularly promising tool here: their split structure of an encoder and decoder allows for multiple decoders to work alongside each other, each targeting a different component of the observed data. In the simplest case which this paper explores, these could aim to extract only two components (the pure signal and background models), but more complicated applications can be explored where different signal components are isolated by separate decoders. This allows for complex signals and backgrounds to be disentangled from each other. 

The proposed model architecture consists of a deep 2D convolutional variational autoencoder model. Autoencoders are a class of neural networks where the model attempts to compress the inputs into a lower dimensional representation (usually referred to as the latent space) before returning the original input as an output. The network is broken down into an encoder which takes the input and compresses it down to the latent space, and the decoder which takes the latent space and decodes it back to the same size as the input, hopefully reconstructing the input (or some modified version of the input) in the process. Variational autoencoders build upon this concept by replacing the simple latent space representation by a probability distribution. This allows the model to learn underlying distributions in data and provide smooth interpolations between data points increasing the networks ability to generalise especially in sparsely or unequally sampled datasets. This work specifically uses deep 2D convolutional variational autoencoders, which simply means instead of a traditional fully connected neural network, the bulk of the model with the exception of the latent space uses convolutional neural network architecture with a depth greater then just a few layers.

In our proposed model, each energy bin in the input data is handled as a separate channel in the encoder/decoder models. The decoder model is designed such that two decoder heads are run in parallel, with each head trained to predict a single cleaned component of the input image. In this particular work, these are the signal of interest and an assumed diffuse background, but the model can easily be modified to target any combination of components. Two versions were explored, the first uses an encoder with 4 convolutional layers and a decoder with 5 layers, while the second uses 8 layers in the encoder and 8 layers in the decoder. Both use a linear variational latent space layer between the encoder and decoder heads. Down sampling is handled with 2D max pooling~\cite{726791} and up sampling with 2D convolutional transposing~\cite{NIPS2012_4824}. No significant performance improvement was identified from the deeper model when handling simpler signal/background combinations but at a significant runtime cost. As a result the shallower model was selected for all testing performed herein.

\newpage
\subsection{Recombination of Network Outputs}

Since the network produces two normalised output models, the relative scale, effectively the signal-to-background ratio, is not natively returned. Instead the signals require recombining and the relative scale recovering by using a two-stage least-squares projection and fitting to the input image. For a simple scenario where the majority of the image is background dominated except for a localised signal this is a trivial fitting problem. However, as we will see later even when this assumption is not perfectly true, such as signal is present across the whole observation or background is dominant across the entire image, it provides a close enough approximation to begin iteratively updating relative scale factors until within a certain numerical tolerance.

To begin with, an initial background estimate is produced by creating a mask of the output signal for spatial and energy bins where the signal is $<1\%$ of the brightest bin, 
we will call this the ``background mask'' ($M_b$) and the inverted mask is referred to as the ``signal mask'' ($M_s$). Using this mask the initial background scale is estimated by solving the first stage least squares fit on the output background ($\mathcal{B}$) to the masked input observation ($\mathcal{O}$). We will call this scale value $\alpha$ and it is determined as follows:
\begin{align}
    \alpha &= \frac{\sum_{i\in M_b}\mathcal{O}_i\cdot \mathcal{B}_i}{\sum_{i\in M_b}\mathcal{B}_i^2}
\end{align}
The signal scale ($\beta$) is then found from the residual ($r$) in the area covered by the signal mask using the second stage of the least square projection:
\begin{align}
    r &= \mathcal{O} - \alpha \cdot \mathcal{B}\\
    \beta &= \frac{\sum_{i\in M_s}r_i\cdot \mathcal{S}_i}{\sum_{i\in M_s}\mathcal{S}_i^2}
\end{align}
These provide us with an initial estimate for $\alpha$ and $\beta$. To improve upon these estimates the second stage can then be iteratively applied to both signal and background using the following:
\begin{align}
    r &= \mathcal{O} - \beta \cdot \mathcal{S}\\
    \alpha &= \frac{\sum_{i\in M_b}\mathcal{r}_i\cdot \mathcal{B}_i}{\sum_{i\in M_b}\mathcal{B}_i^2}\\
    r &= \mathcal{O} - \alpha \cdot \mathcal{B}\\
    \beta &= \frac{\sum_{i\in M_s}r_i\cdot \mathcal{S}_i}{\sum_{i\in M_s}\mathcal{S}_i^2}
\end{align}
This is done until convergence is reached (defined as the change in $\alpha$ and $\beta$ being $<10^{-20}$) or a maximum number of iterations (20) is reached.

Consideration for future work is to include a regression output from the model itself estimating the relative scale factor between the two models. This would eliminate the need for a post-processing step and be vastly more robust for scenarios with no background dominated region, poor and/or very low S/B reconstructions or scenarios with degenerate solutions for recombination.

\subsection{Data Sample Training}
Training was performed on a dataset of 64000 unique observations split into 1000 batches of 64. For the complex toy model scenario, this number was doubled to 128000 unique observations in 2000 batches (see later for more details). The training aims to minimise the loss function defined as the sum of the mean squared errors (MSE) between each target component and the reconstructed image of each component. This loss can optionally be extended with KL-divergence~\cite{Kullback:1951zyt} and sparsity penalties~\cite{Goodfellow-et-al-2016}. These can theoretically help with reconstruction of non-trivial signal and background forms by helping the VAE to understand the underlying distributions of observations better and helping to reduce ``salt and pepper'' noise~\cite{GonzalezWoods2018}
on sparse signal and/or background observations (\textit{i.e.,} where there are no gamma-rays in most of the image), respectively. This work only uses uniformly sampled spaces of observation with no sparse observations and hence both are kept at zero throughout this work, but they are mentioned here for their potential use in future applications.

Each model is allowed to train for 100 epochs before training is terminated. The loss function is observed to monotonically decrease until approximately epoch 70-85 (depending on the scenario), and then slowly drop from there until minimal or no change occurs indicating convergence has been reached. Further test runs extended this out to 1000 epochs to verify that no further decrease was observed, and these confirmed that the model reaches a stable solution by 100 epochs every time. Approximately 10\% of models diverge early and fail to train, which is readily detectable after one to two epochs (due to rapid divergences to infinity triggering the loss function to become nan). In these cases, the training is aborted and restarted. Optimisation is performed with an Adam optimiser~\cite{kingma2017adammethodstochasticoptimization} using default settings~\cite{AdamSettings} 
and a learning rate of 0.00001. Training generally takes between 35 and 50 seconds per epoch or around 1 to 1.5 hours (including startup time) to train a single model. This is doubled for the complex scenario with twice the data. A full ensemble of 10 models can therefore be trained in 10-to-15 hours by training in series, although training in parallel can significantly reduce this time depending on hardware capacity. Training was performed on an NVIDIA RTX3090 accompanied by an AMD 5950x 16 core/32 thread CPU and 128GB memory.
\begin{table}[h!]
\centering
\begin{tabular}{l|c}
\hline
\hline
\textbf{Networks} & $\mathbf{S_s, S_b}$\\
\hline
Input dimension and coordinate & $3,\ (\vec{x}, e)$\\
Number of layers & 10 (4 encoder, 5 decoder, 1 latent)\\
Down sampling & max-pool\\
Up sampling & 2D convolutional transpose\\
Convolutional layer Activation function & ReLU\\
Final Activation function & Sigmoid\\
Output dimension and coordinate & $4,\ (n, \vec{x}, e)$ 
\\
Latent space size & 64\\
\hline
\hline
\end{tabular}
\caption{Hyperparameter settings in the VAE-based network architecture used in this work. All the VAE are trained using the Adam optimiser with a learning rate of 10$^{-5}$. $n$ stands for the number of target components.}
\end{table}

\newpage
A single VAE is a powerful tool, however due to the nature of signal-background separation, variation in the output from separately trained methods is to be expected. This is because each VAE converges to a similar but slightly different solution. In ``simple'' to reconstruct scenarios the agreement between these solutions will be extremely close, but in more complex scenarios, such as extremely low 
signal-to-background ratio cases, the agreement will be a lot lower. To improve the performance, multiple VAEs can be utilised in parallel with the outputs averaged in an approach referred to as an ensemble model. A significant advantage of this approach outside of the averaging over multiple solutions is that it allows for a confidence interval on the results to be established. This provides a useful indicator of how well an event is reconstructed and can be used to create an error band on the reconstructed distributions. Close agreement between models indicates high reconstruction confidence, while a wide error band indicates low model confidence.
\begin{figure}
    \centering
    \includegraphics[width=1.0\linewidth]{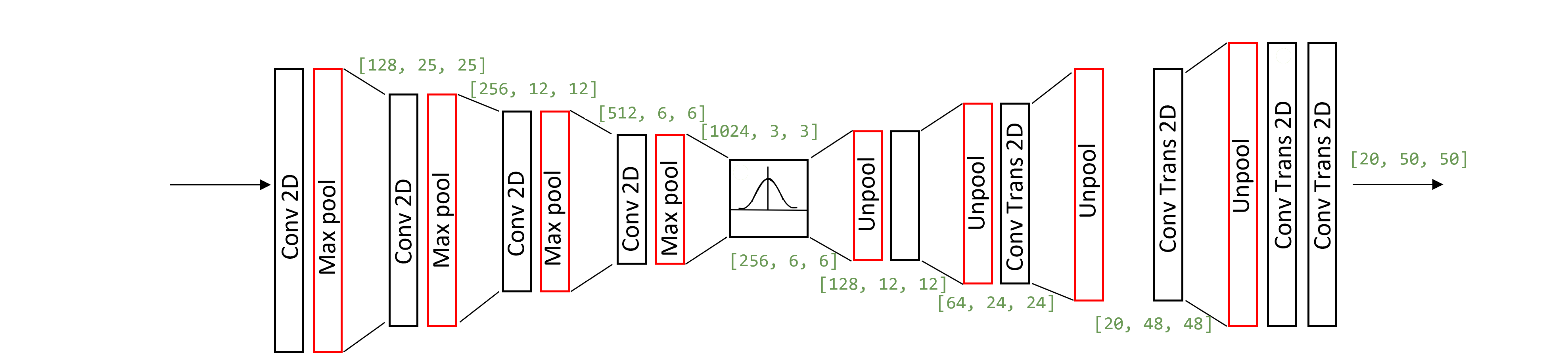}
    \caption{Diagram of the convolutional autoencoder architecture used in this work. The encoder (left) consists of four 2D convolutional layers, each followed by max-pooling. The resulting latent representation is then decoded by a symmetric sequence of unpooling and transposed-convolution layers, reconstructing the original 50×50 image. Green labels indicate the tensor dimensions [channels,height,width] at each stage of the network.}
        \label{fig:network small}
\end{figure}

\section{Data Format and Data Sets}
\label{sec:dataset}
In order to validate the architecture described above, we have applied to a variety of test models of increasing complexity, culminating in a realistic analysis of H.E.S.S. data. 
\subsection{Formatting Data Sets}
The data are arranged in a 3D data cube 
assuming 50$\times$50 spatial bins with a bin width of 0.1$^\circ$ and 20 logarithmically-spaced energy bins
covering an energy range of 0.1 to 10 TeV. A Poisson realisation of the background and signal components is taken prior to combination to introduce a realistic amount of statistical noise to the observation. Hence the total model consists of the expression given in Eq.~(\ref{eq:ftot}). Separate 3D data arrays are generated for the signal and background components which, in different tests, are combined with different weightings to achieve a desired 
signal-to-background ratio on the brightest spatial pixel, hereafter referred as to SB. 

\newpage
\subsection{Toy Data Set}
Our first model is a simple toy model that utilises simple spatial models and power law spectral models for both the signal and background components, with the parameters designed to mimic typical H.E.S.S. observations. 
The individual spatial and energy 
count models are given by:
\begin{align*}
    \begin{cases}
       S_{s_i}(\overrightarrow{x}) &= C_s\exp \left(\dfrac{\overrightarrow{x_i}-\overrightarrow{x}_0}{2r_i^2}\right)^2\, ,\\
       E_{s_i}(e) &= \alpha_s \left(\dfrac{e}{E_0}\right)^{-\Gamma_{s_i}}\times T \times A(e)\, ,\\
       S_b(\overrightarrow{x}) &= C_b\, ,\\
       E_b(e) &= \alpha_b \left(\dfrac{e}{E_0}\right)^{-\Gamma_b}\times T \times A(e)\, ,\\
    \end{cases}
\end{align*}
where $\Gamma_{s_i}$ and $\Gamma_{b}$ are independent randomly generated spectral indexes in the  range $[1.5,3]$ for the signal and background components, $A(e)$ is the effective area for a typical IACT like H.E.S.S. in a given energy bin and $T$ is the assumed observation time (set arbitrarily to 1800 seconds). The normalisation parameters $\alpha_s$ and $\alpha_b$ are set such that $\int E_s\,dE_s=\int E_b\,dE_b=1$. Finally $C_s$ and $C_b$ are set such that the signal-to-background ratio of the brightest pixel is at a target value, SB, and such that the average particle count per bin is 10. This is given by:
\begin{align*}
    C_s = \dfrac{10\, C}{\overline{CF_s + F_b}}\, , \quad
    C_b = \dfrac{10}{\overline{CF_s + F_b}}     \, , \quad
    C = SB \times \dfrac{\max(F_b)}{\max(F_s)}\, ,
\end{align*}
where $C$ is an intermediate value used to obtain the targeted signal-to-background ratio.

 A full dataset as training sample is then generated from this above model, by uniformly sampling from the parameter space of inputs into the model. Each sample can therefore be represented by the vector $[\overrightarrow{x_i}, r_i, \Gamma_{s_i}, \Gamma_b, SB]$ which can be thought of as a scaled and shifted unit vector in a hyper cube. To achieve good performance, this sampling needs to be fine enough to contain a sufficient coverage of the hypercube. For the simple case (\textit{i.e.}, one source in the image), we use 64,000 samples averaging to 6 samples per axis providing a relatively sparse but still good amount of data. For the complex case where the number of sources varies between 0 and 4, we utilise twice the training data however we only have 2.6 samples per axis (and noting that these samples are biased towards zero on the position axes due to the variable number of sources) meaning the complex case is significantly under sampled. However, this provides a good chance to explore the limits of our approach in sparsely sampled datasets.

An important thing to note about this toy data model is that it is a highly simplified scenario with a very basic Gaussian shape and power-law spectrum. It is therefore a rough approximation of a potential source that is chosen for its ease of generation for testing purposes rather than for its realism. In this model, a single observation can be generated in a few hundred milliseconds, with a full data set made ready in a couple of hours. A more realistic dataset would require orders of magnitude more time to run full simulations of, limiting the ability for the data to be adjusted and the model performance to be explored. As a result of this, most ``real'' observations fall outside of the range of scenarios modelled by the data mostly due to not having the simple structure assumed by this model, but also potentially due to simply not being in the range of parameters sampled for the toy model.

\subsection{Mock Dark Matter Data In The Galactic Centre}
A second model provides mock H.E.S.S.-like observations of the Galactic centre based closely on real observations taken by the H.E.S.S. Inner Galaxy Survey~\cite{HESS:2022ygk}. This took a high exposure observation of VHE gamma-rays in the inner few degrees of the Milky Way halo, with over 546 hours of observation in total. A mock dark matter model implements a 2-component mixture of a VHE gamma-ray signal induced by the annihilation of Majorana dark matter particles in the Milky-Way DM halo and a residual background component. The energy-differential residual background flux from H.E.S.S. is extracted from Ref.~\cite{HESS:2022ygk}, which is assumed to be spatially isotropic. This assumption is justified, with no significant spatial dependence being found down to 1\% level. We express the signal component of the mixture model using the following energy-differential flux of gamma-rays in a solid angle $\Delta\Omega$ expected from the pair-annihilation of Majorana DM particles of mass $m_{MD}$ and in a DM halo of DM density $\rho$ given by:
\begin{equation}
\label{eq:dmflux}
\frac{d \Phi}{d E}(E) =
    \frac{\langle \sigma v \rangle}{8\pi m_{\rm DM}^2}\sum_f  {\rm BR}_f \frac{d N_f}{d E}(E) \, J(\Delta\Omega) \quad {\rm with} \quad
J(\Delta\Omega) = \int_{\Delta\Omega} d \Omega \int_{\rm los}  d s\, \rho^2(s[r,\theta]) \, .
\end{equation}
The expected signal counts are computed for the self-annihilation of DM of 1 TeV mass in the $W^+W^-$ channel using an annihilation cross section of 4$\times$10$^{-25}$ cm$^3$s$^{-1}$, and the Einasto parametrisation of the DM density profile using profile parameters taken from Ref.~\cite{HESS:2022ygk}. We assume a constant time exposure of 500 hours with an assumed Gaussian energy resolution with width $\sigma$/E = 10\% and a SB value 
between 0.25 and 2 
with events binned in 30 logarithmically-spaced bins between 0.1 and 100 TeV, and 39 bins in Galactic longitude in and latitude between -10\textdegree and +10\textdegree. Due to the complex nature of the simulation, only a single simulation run is performed (referred to as the template event). From here the training sample is obtained from Poisson realisations performed on the template and the signal-to-background ratio is adjusted by rescaling the realisation of the signal template. The real observation is excluded from the training dataset to allow for a fair test of the model's performance.

\subsection{H.E.S.S. Public Data Release}

The final dataset used for testing takes the form of real observations of the Crab Nebula and the pulsar wind nebula MSH 15-52 from the H.E.S.S. public data release~\cite{HESSpublicrelease}. Using a field of view of 1.5 degrees centred on the source position in the energy range from 0.1 to 10 TeV with binning kept the same as used in previous data sets, the histograms are constructed by stacking observational runs and converting the resulting event list into a overall 3D histogram. The SB value 
is estimated to be of around 20 for Crab and 0.8 for MSH 15-52. As discussed above, the toy model fails to cover these real world scenarios with non-Gaussian spatial structure and non-trivial energy spectrum. To provide a broad enough training set, the H.E.S.S. public observations were used as a template to generate Crab and MSH 15-52 like observations. The energy spectrum of the signal and signal-to-background ratio are modified before taking a Poisson realisation of the templates. The SB value is randomised within a range of 15\%-200\% of the original, while the spectrum is modified by stretching or compressing the spectral shape and providing a shift in spectral peak. This allows for the production of a broad but well sampled dataset covering similar observations to the object of interest. Importantly, like with the DM scenario, the real observation is explicitly excluded from the training dataset to provide a fair test of the model.
Unlike the dark matter scenario, the template model for both H.E.S.S. observations contains large noise spikes away from the source significantly above the Poisson statistical noise. To help prevent model overfitting, a random Gaussian smoothing is applied (radius $\in [0, 5]$) to the template prior to the Poisson realisation of the template being taken.

\section{Results}
\label{sec:results}
\subsection{Toy Model}
We will first present the results of the model's application to the simple toy model and then explore the extension to more complex scenarios. The toy model allows for precise control over all parameters of the sample observation, including the signal to noise ratio, position and shape of the count events, and the signal and background spectral indices. With known ground truth models available, the performance can be directly evaluated by comparing reconstructed distributions with the ground truth. To begin with, we will make use of a fairly simple scenario to act as a baseline toy scenario, in this case a SB value of 0.68 with a radius of 0.3\degree~using a signal and background spectral indices of 2.2 and 2.4, respectively.
\begin{figure}[htbp]
    \plotfromdir{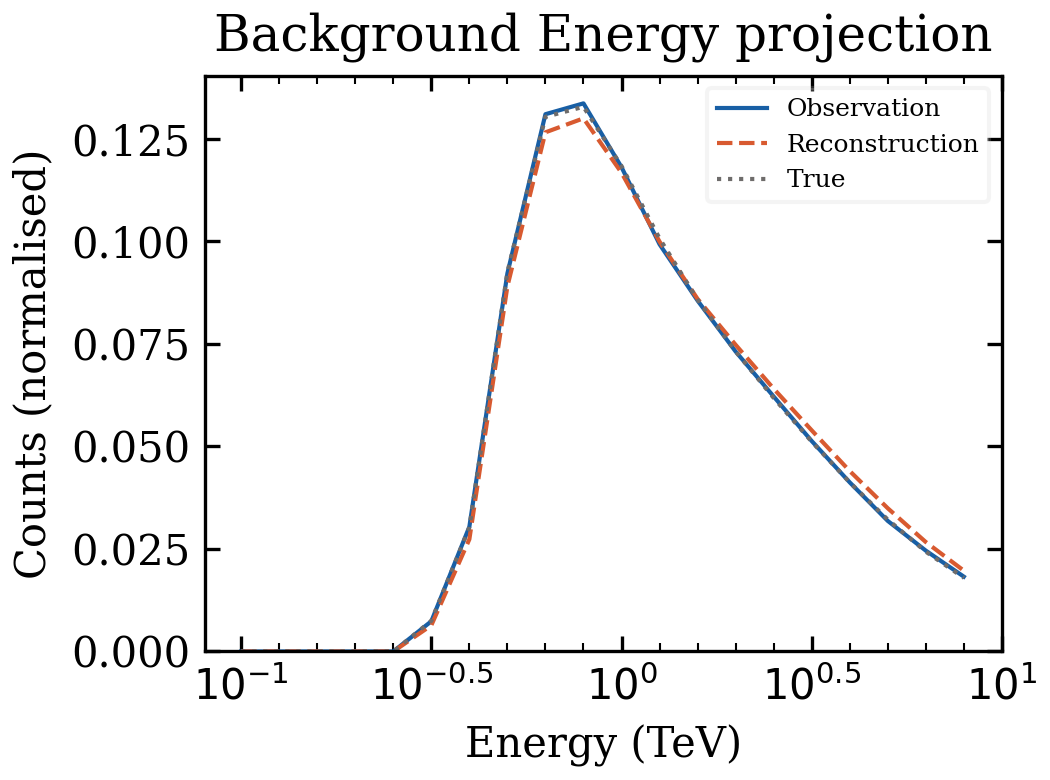}
    \caption{Comparison of the neural network output with the assumed observed dataset and ground truth for the baseline toy model case: an SB value of 0.6 with a radius of 0.3$^\circ$ for the spatial component, spectral indices of 2.2 and 2.4 for the signal and background energy power-law models, respectively.  
    \textit{Top panels:} spatial distributions of the true (left) and reconstructed model components (right). The reconstructed components correspond to the denoised output. 
    \textit{Middle panels:} The left and middle plots show the normalised projected spatial distributions for the signal and  background components, given as 1D distributions in $x$ and $y$. The left and right axes label the signal and background components, respectively. Shown for comparison are the true distributions for the assumed signal and background models (green dotted lines), the simulated observation which includes Poisson noise (blue solid lines) and the reconstructed denoised model output (red dashed lines). The right panel shows the signal energy distribution, with the same convention on the individual lines.
     \textit{Bottom panels:} The left and right panels show the normalised projected spatial distributions summed over the signal and  background components. The right panel shows the background energy distribution.} 
    \label{fig:single train}
\end{figure}

In the top left panel of Fig.~\ref{fig:single train}, we show the assumed observed gamma ray image in the spatial plane, including Poisson noise. The output of our reconstruction is a denoised and precise estimate of the individual signal and background components that were present in the original image, shown in the spatial plane in the top right of Fig.~\ref{fig:single train}. The 1D marginalised spatial distributions of the $x$ and $y$ components of the image are shown in the middle row for the individual signal and background components, where we show the original true distribution from which our noisy signal distribution is sampled, the assumed observation which includes Poisson noise, and the output of our reconstruction. The reconstructed denoised components closely match the ground truth distribution as should be expected. In the bottom panel, we show the total observation noting that, since the model outputs are not necessarily scaled correctly, they need to be corrected when recombining to match the input observation. However, since each component can be trivially normalised based on the construction of the toy model, the shape and relative difference in the models can be easily explored and analysed. Figure~\ref{fig:single train} also shows the spectral distribution for the true, noised (observed) and reconstructed cases, again demonstrating close agreement between the true and reconstructed distributions.

To better 
quantify the performance, we utilise the weighted relative difference between the predicted $F_{\rm  out}$ and the ground truth $F_{\rm true}$ as follows:
\begin{equation}
\label{eq:dr}
    D_{\rm R}=\frac{\sum|F_{\rm out}-F_{\rm true}|}{\sum|F_{\rm true}|}\, ,
\end{equation}
where the sum symbol corresponds to a discrete sum of the field values over the bins of the 3D coordinate space.
This metric is adequate for understanding the overall quality of the reconstruction, but is fundamentally sensitive to the noise in the original observation. The KL-divergence can also be calculated to provide an estimate of how well the model is recreating the underlying distribution without noise. 
While the smallest values of the KL-divergence provide the best agreement between the input and output model, values below 0.01
are interpreted as strong agreement between the two distributions.
The $D_{R}$ and KL-divergence values provide an estimate of how well the model has reconstructed the input, the noise and the underlying distribution. For the above scenario we get a overall $D_{R}$ of 3.55\% and KL-divergence value of 0.0012. The signal/background and spatial/energy models can be further subdivided with the energy spectrum having a $D_{R}$ of 6.50\% and KL-divergence value of 0.0045 for the signal and likewise 3.20\% and 0.00093 for the background and the spatial reconstructions ($X/Y$) having $D_{R}$ and KL-divergence value of 2.07\%/10.00\% and 0.0031/0.015 for signal and background having 0.33\%/0.30\% and 9.1$\times$10$^{-6}$/8.4$\times$10$^{-6}$, 
as summarised in Tab.~\ref{tab:Simple}.
This tells us that the model is performing well, especially reconstructing the underlying distribution while ignoring the noise in the observation. 
\begin{table}[htbp]
    \centering
    \begin{tabular}{c|c|c}
        \hline        \hline
         Value& $D_{\rm R}$& KL \\
         \hline
         Overall& 3.55\%& 0.0012\\
         Energy spectrum signal& 6.50\%& 0.0045\\
         Energy spectrum background& 3.20\%& 0.00093\\
         Spatial X signal& 2.07\%& 0.0031\\
         Spatial Y signal& 10.00\%& 0.015\\
         Spatial X background& 0.33\%& 0.0000091\\
         Spatial Y background& 0.30\%& 0.0000084\\
         \hline        \hline
    \end{tabular}
    \caption{Results on the relative difference $D_{\rm R}$ and KL-divergence for the signal and background components in the baseline toy model scenario: an SB value of 0.6 with a radius of 0.3$^\circ$ for the spatial component, spectral indices of 2.2 and 2.4 for the signal and background energy power-law models, respectively.  }
    \label{tab:Simple}
\end{table}

\subsection{Toy Model With Ensemble}
\begin{figure}[htbp!]
    \plotfromdir{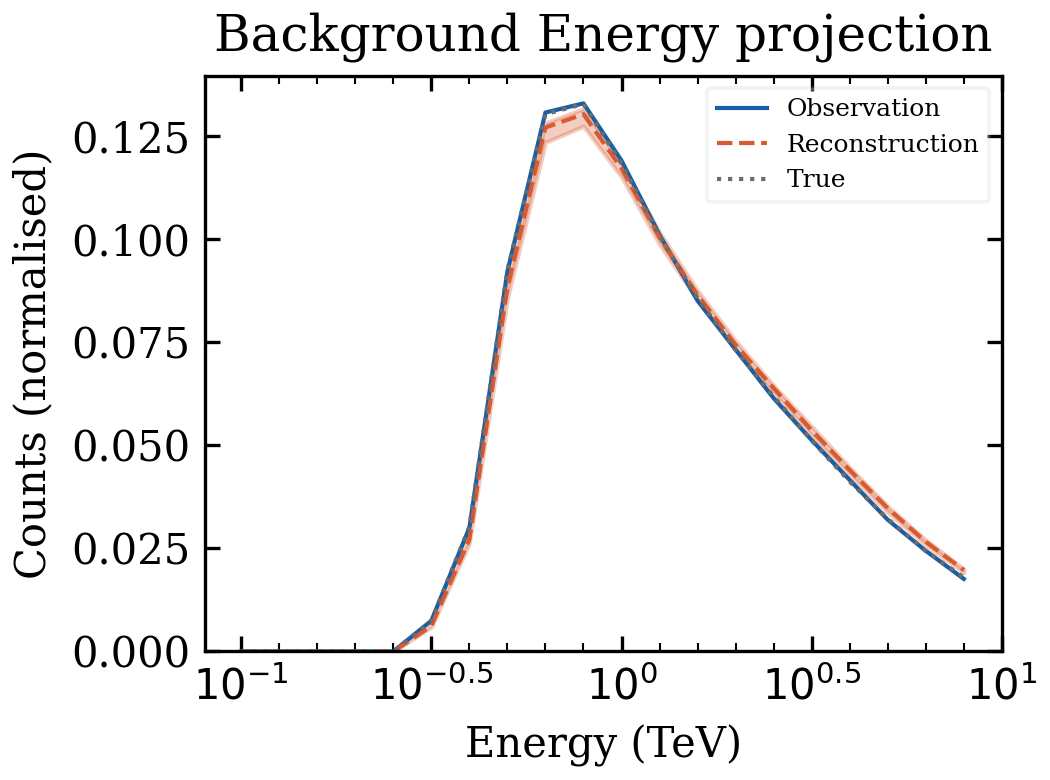}
 \caption{Same as Figure~\ref{fig:single train}, but for an ensemble of 10 trained models with different initial conditions for the neural network parameters. The bands in the middle and bottom panels provide the 68\% confidence band of the model outputs for a given component.}
    \label{fig:multi train}
\end{figure}

One observation that can be made of the model, is that while the model outputs are stable, retraining the model does produce significant differences in the resultant output. To remedy this spread and improve overall performance, an ensemble model can be utilised using 10 models in parallel.
In each model the initial  conditions of the NN  are different, \textit{i.e.}, the values of the weights and biases are randomly modified following a normal distribution.
Although more computationally expensive, the overall performance is greatly increased as is seen by the mean in Fig.~\ref{fig:multi train}. Additionally, we gain the ability to provide an uncertainty estimate on the results by producing a 68\% containment band on the models output. We will again use a very simple scenario to test the performance of the model using the toy model with a SB value of 0.73, a radius of 0.3\degree using a signal spectral index of 2.2 and background of 2.4.
As shown in Fig.~\ref{fig:multi train} the ensemble model is able to produce a better result then the single model approach. Additionally, the width of the error band gives a strong indicator of the models confidence in the results, with the band being very narrow on events with a high confidence of reconstruction, and widening as the observation becomes more difficult to reconstruct for the model. 
Like with the previous scenario, the relative difference and KL-divergence can be calculated, and these are presented in Tab.~\ref{tab:Simple x10}.
\begin{table}[htbp]
    \centering
    \begin{tabular}{c|c|c}
        \hline\hline
         Value& $D_{R}$& KL \\
         \hline
         Overall& 3.65\%& 0.0013\\
         Energy spectrum signal& 3.87\%& 0.0010\\
         Energy spectrum background& 3.17\%& 0.00096\\
         Spatial X signal& 8.48\%& 0.0096\\
         Spatial Y signal& 6.20\%& 0.0054\\
         Spatial X background& 0.89\%& 0.000077\\
         Spatial Y background& 0.85\%& 0.000069\\
         \hline\hline
    \end{tabular}
    \caption{Ensemble results on the relative difference $D_R$ and KL-divergence for the mean signal and background components in the baseline toy model scenario from an ensemble output based on 10 model runs.}
    \label{tab:Simple x10}
\end{table}
The model again recreates the underlying model well while ignoring the noise in the observation. Again the overall reconstruction is precise favouring a smooth noiseless output from the model rather than a noisy one.

\newpage
\subsubsection{Low Signal-to-Background Ratio Scenario}
\begin{figure}[htbp!]
    \plotfromdir{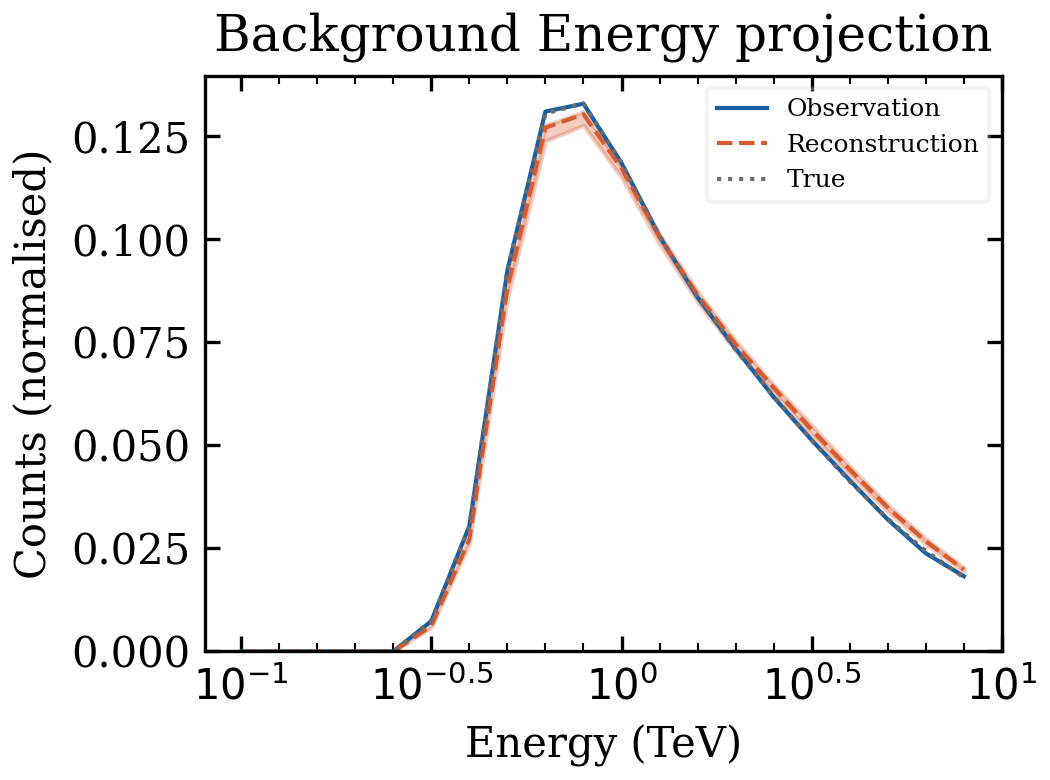}
    \caption{Comparison of the neural network output with the assumed observed data and ground truth for an ensemble of 10 model runs for the toy model case with a modified spatial distribution (SB value of 0.17, radius of 0.3$^\circ$). The signal and background spectral components are unchanged with respect to Figure~\ref{fig:multi train}.   
}
    \label{fig:low SB}
\end{figure}

With the reconstruction proving effective on the toy model under ``normal'' circumstances, we can now experiment with pushing the model to its limits. To test this we create a scenario with  
low signal-to-background ratio, \textit{i.e.}, ~10\% in the brightest pixel. The exact parameters of this scenario are given by an SB value of 0.1, a radius of 0.3\degree, a signal spectral index of 2.2 and a background spectral index of 2.4. One important thing to note from 
Figure~\ref{fig:low SB} is that the noise in the input is sufficiently large enough to create significant deviation from the smooth input Gaussian. As a result, performance statistics are expected to perform significantly worse then in previous examples. The relative difference and KL-divergence are presented in Tab.~\ref{tab:low SB}.
\begin{table}[htbp]
    \centering
    \begin{tabular}{c|c|c}
        \hline     \hline
         Value& $D_{R}$& KL \\
         \hline
         Overall& 4.20\%& 0.0017\\
         Energy spectrum signal& 12.90\%& 0.011\\
         Energy spectrum background& 3.30\%& 0.0010\\
         Spatial X signal& 52.46\%& 0.28\\
         Spatial Y signal& 56.13\%& 0.27\\
         Spatial X background& 1.31\%& 0.00019\\
         Spatial Y background& 1.19\%& 0.00016\\
         \hline     \hline
    \end{tabular}
    \caption{
    Ensemble results on the relative difference $D_R$ and KL-divergence for the mean signal and background components in the toy model scenario from  an  ensemble output based on 10 model runs: an SB value of 0.17 with a radius of 0.3$^\circ$ for the spatial component, spectral indices of 2.2 and 2.4 for the signal and background energy power-law models, respectively.
    }
    \label{tab:low SB}
\end{table}

\newpage
Looking at the outputs, we can see that the model is faithfully reconstructing the inputs within error. However, as expected we can see that the relative difference and KL-divergence are significantly worse then the previous examples due to the large amount of statistical noise present. Despite this, the results show the model was still able to perform well under extreme conditions.

\begin{figure}[htbp!]
    \plotfromdir{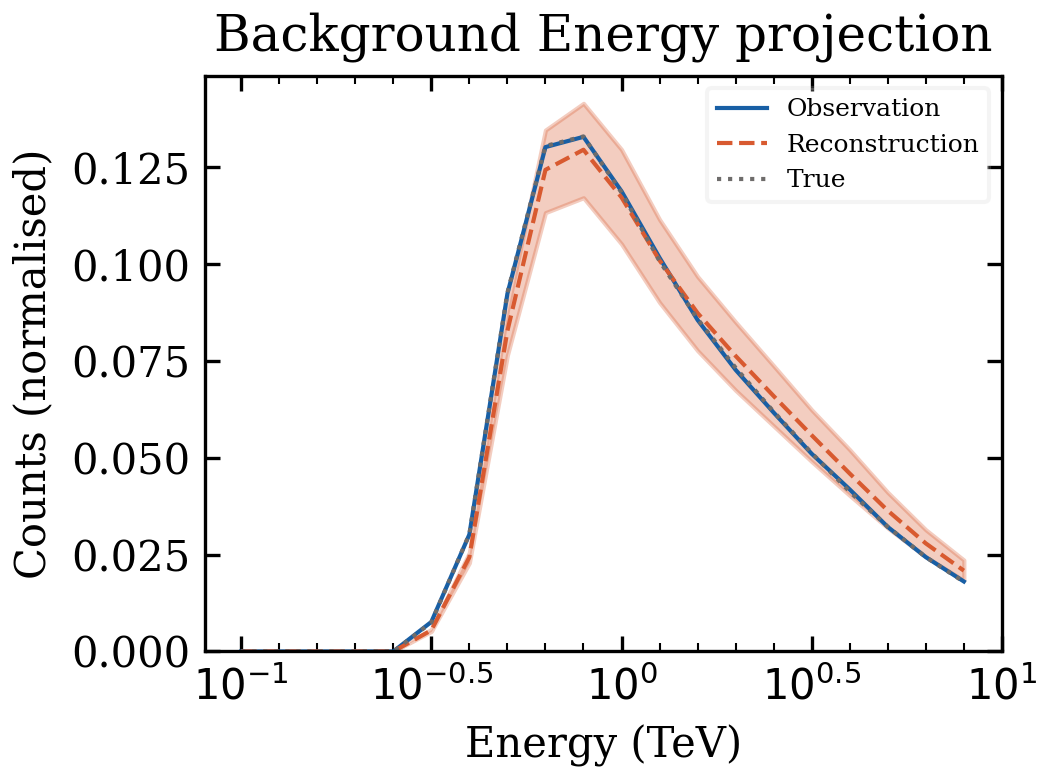}
    \caption{
Ensemble neural network output model for the model case with three signal and one background components. 
The input model is composed of 4 signal sources with the parameters
(x position, y position, radius, spectral index) given by (1.22, 1.5, 0.4, 2.2), (1.52, 2.275, 0.3,378
2.2), (2.39, 1.26, 0.4, 2.1) and (4.0, 4.0, 0.25, 2.3), respectively. 
The overall SB value is 0.028. The background spectral index is 2.4.
10 models run with different initial conditions are used. The bands in the middle and bottom panels provide the 68\% confidence band of the model outputs for a given component.  
    }
    \label{fig:Complex}
\end{figure}

\subsubsection{Multiple Component Scenario}
The final toy model scenario explored in this work is the ``complex'' scenario. In this example, the model is retrained with a variable number of sources between zero and four sources per observation. The input model 
is composed of 4 signal sources with the parameters (x position, y position, radius, spectral index) given by (1.22, 1.5, 0.4, 2.2), (1.52, 2.275, 0.3, 2.2), (2.39, 1.26, 0.4, 2.1) and (4.0, 4.0, 0.25, 2.3), respectively.  The background spectral index is 2.4 , and the overall SB value is 0.7.
This leads to complex observations with 
extended overlapping source-like structures. Figure~\ref{fig:Complex} shows the model reconstruction of a combination of 4 sources, with 3 in very close proximity and two  with significant overlap. As can be seen the model is able to still provide a good reconstruction of the overall shape, however it struggles to perfectly match the ground truth, especially in the trough between the split peak in the x axis. The full summary statistics can be found in Tab.~\ref{tab:Complex}.
\begin{table}[htbp]
    \centering
    \begin{tabular}{c|c|c}
        \hline        \hline
         Value& $D_{R}$& KL \\
         \hline
         Overall& 7.42\%& 0.0049\\
         Energy spectrum signal& 4.84\%& 0.0021\\
         Energy spectrum background& 5.72\%& 0.0031\\
         Spatial X signal& 16.87\%& 0.056\\
         Spatial Y signal& 13.58\%& 0.042\\
         Spatial X background& 3.33\%& 0.00094\\
         Spatial Y background& 2.52\%& 0.00057\\
         \hline  \hline
    \end{tabular}
    \caption{Ensemble results on the relative difference $D_R$ and KL-divergence for the mean components in the four-component-signal scenario for 10 runs. The number of signal components is let free in the output model.}
    \label{tab:Complex}
\end{table}

\subsection{Mock Dark Matter}
We now explore the model's performance on the mock dark matter data set. This is a significantly more complicated example with the DM signal covering the entire field of view with a low signal-to-background ratio across the entire field of observation. While it was expected that significant tweaks to the model configuration would be required, we found that only minor tweaks were necessary to the network structure, \textit{i.e.}, the input and output layer sizes as well as input and output sizes of the decoders to accommodate the slightly different data size.

\begin{figure}[htbp!]
    \plotfromdir{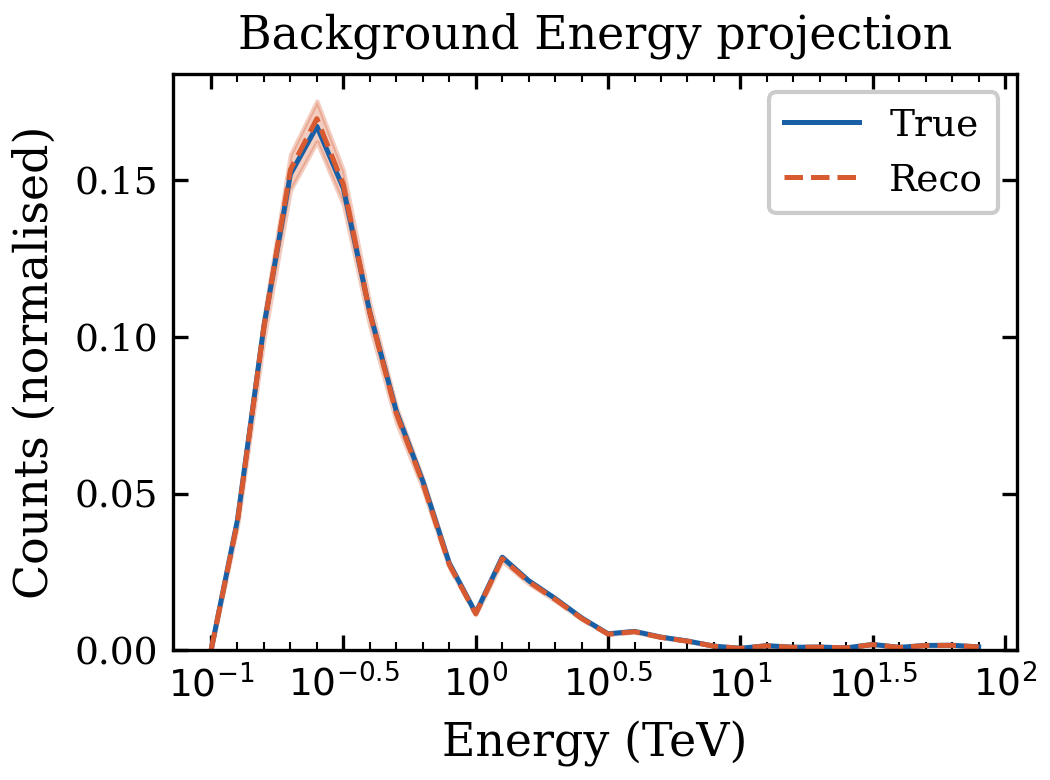}
    \caption{Comparison of the neural network output for the mock dark matter scenario for an ensemble of 10 model runs. The
models run with different initial conditions are used. The bands provide 68\% of the model
outputs (middle and bottom panels).}
    \label{fig:DM}
\end{figure}

The performance statistics for the mock DM model are seen in Tab.~\ref{tab:DM}. The error band shown in
Fig.~\ref{fig:DM} is very narrow and the mean value is close to the ground truth, indicating consistent performance. Most importantly, the model has accurately detected the $W^+W^-$ annihilation channel ``bump'' near the DM mass in the energy spectrum. This indicates that the model is able to reliably detect DM features and could potentially identify multiple DM annihilation channels if present. Surprisingly, despite there being no region without significant signal present, the recombination was able to successfully obtain the SB and recombine the signal and background into the observation, we will see similar results on the real data in the next section.
It is worth noting that the dataset only covers a single annihilation channel and DM mass, so the generalised performance to other DM channels based on this dataset remains unknown. However, extending the dataset to cover other possibilities would be a trivial extension in future work, and likewise more realistic models can be used to further improve real world performance.
\begin{table}[htbp]
    \centering
    \begin{tabular}{c|c|c}
         \hline\hline
         Value& $D_{R}$& KL \\
         \hline\hline
         Overall& 3.16\%& 0.0012\\
         Energy spectrum signal& 1.26\%& 0.00017\\
         Energy spectrum background& 0.95\%& 0.000056\\
         Spatial X signal& 1.14\%& 0.00012\\
         Spatial Y signal& 1.23\%& 0.00013\\
         Spatial X background& 0.75\%& 0.000037\\
         Spatial Y background& 0.57\%& 0.000025\\
         \hline\hline
    \end{tabular}
    \caption{Ensemble
    results on the relative difference $D_R$ and KL-divergence for the mean signal and background components in the mock dark matter model scenario from  an  ensemble output based on 10 model runs. 
    }
    \label{tab:DM}
\end{table}

\subsection{Real Data}
Finally, this work considers the two real world examples discussed earlier. Being real data, there is no known ground truth for these observations, so instead the model output will be compared with the 
spatial and energy spectrum models for the two objects obtained from standard extraction methods. As a result constant offsets from zero can occur which require a simple post processing step to correct for in the signal plots. The recombination process naturally accounts for this in the total observation however. Energy spectra are estimated with the \textit{Multiple OFF} method~\cite{Berge:2006ae} while the spatial distribution makes use of the \textit{Ring Background} method~\cite{Berge:2006ae}.

\subsubsection{Crab Nebula}
\begin{figure}[htbp!]
    \plotfromdir{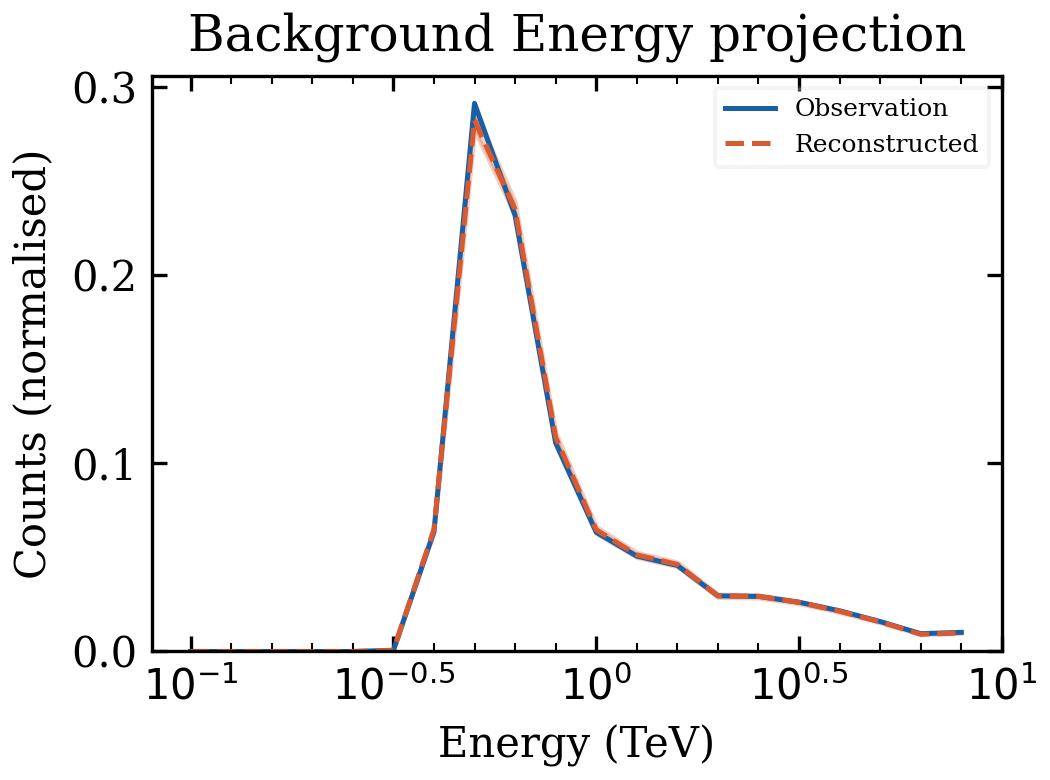}
    \caption{
Comparison of the neural network output for the Crab observed data for an ensemble of 10 model runs. The
models run with different initial conditions are used. The bands provide 68\% of the model
outputs (middle and bottom panels).}
    \label{fig:Crab}
\end{figure}

Figure~\ref{fig:Crab} shows the ensemble model reconstruction 
for the Crab nebula dataset using 10 models. As before, for each model the initial conditions of the NN are different.
The reconstruction output provides a denoised and precise estimate
of the individual signal and background components, respectively. The spatial and spectral output shapes show little scatter from one model to the other as shown by the 68\% containment band on the models output. The relative difference and KL-divergence values  are given in Tab.~\ref{tab:Crab}. The model performs quite well and is able to reliably reconstruct the source and background components.

\begin{table}[htbp]
    \centering
    \begin{tabular}{c|c|c}
        \hline\hline
         Value& $D_{R}$& KL \\
         \hline\hline
         Overall& 31.74\%&  0.10\\
         Energy spectrum signal& 7.46\%& 0.016\\
         Energy spectrum background& 2.03\%& 0.00026\\
         Spatial X signal& 22.67\%& 0.071\\
         Spatial Y signal& 20.92\%& 0.054\\
         Spatial X background& 1.15\%& 0.000099\\
         Spatial Y background& 0.95\%& 0.000082\\
         \hline\hline
    \end{tabular}
    \caption{Results on the relative difference $D_{R}$ and KL-divergence for the mean signal and background components in the Crab nebula data from an ensemble output based on 10 model runs.}
    \label{tab:Crab}
\end{table}

\newpage
\subsubsection{MSH 15-52}
\begin{figure}[htbp!]
    \plotfromdir{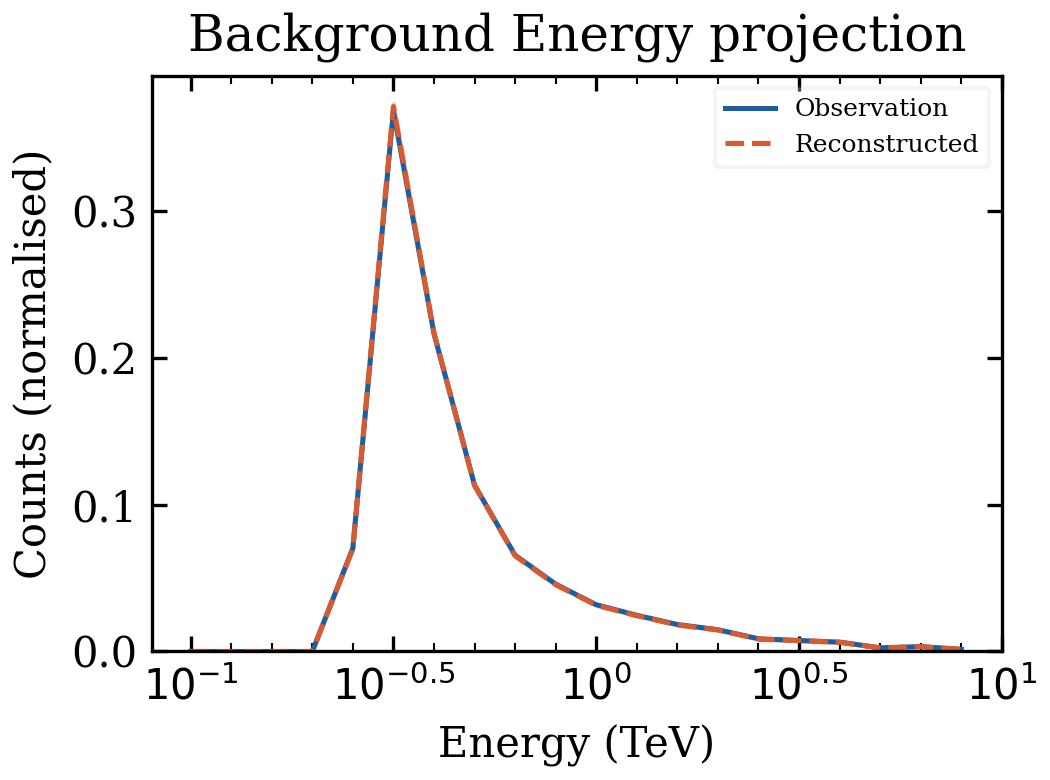}
    \caption{
Comparison of the neural network output for the MSH 15-52 observed data for an ensemble of 10 model runs. The bands provide 68\% of the model outputs in the middle and bottom panels.
    }
    \label{fig:MSH}
\end{figure}

Figure~\ref{fig:MSH} shows the ensemble model reconstruction 
for the MSH 15-52 dataset using 10 models. The model output is able to precisely reconstruct the spatial shape of the source together with the spatial variation in the field of view. The spectral shapes are also reliably reconstructed. The 68\% containment bands on the model output show little scatter in the spatial and spectral output shapes. Table~\ref{tab:MSH} gives the relative difference and KL-divergence values.
For the more complex case of MSH 15-52, where the SB value is lower than for the Crab, and the signal source is spatially extended,  the model performs well and is able to faithfully reconstruct the source and background components. 

\begin{table}[htbp]
    \centering
    \begin{tabular}{c|c|c}
        \hline\hline
         Value& $D_{R}$& KL \\
         \hline\hline
         Overall& 11.74\%&  0.021\\
         Energy spectrum signal& 10.12\%& 0.012\\
         Energy spectrum background& 0.67\%& 0.000037\\
         Spatial X signal& 25.87\%& 0.071\\
         Spatial Y signal& 20.89\%& 0.051\\
         Spatial X background& 0.61\%& 0.000031\\
         Spatial Y background& 0.75\%& 0.000036\\
         \hline\hline
    \end{tabular}
    \caption{Ensemble results on the relative difference DR and KL-divergence for the mean signal and
background components for MSH 15-52 data from an ensemble output based on 10 model runs.}
    \label{tab:MSH}
\end{table}

\section{Conclusions}
\label{sec:discussion}
We sought to solve the problem of signal-background separation in noisy and potentially low signal-background ratio scenarios with a solution that is able to provide accurate estimates of spectral and spatial shapes  of both signal and background, in a non-parametric way, including the ability to provide an estimate of the model uncertainty sometimes coined as the epistemic uncertainty. To achieve this we built on past work using NN methods to estimate the signal to background distributions and extended the approach to use an ensemble of VAEs to completely reconstruct the signal and background at a pixel by pixel level while also removing the aleatoric noise from the inputs. This method proved to be very capable, especially in low signal-background ratio scenarios providing a reliable reconstruction and good error estimate. This method improved on previous works, allowing for the model to be extensible to a much wider range of scenarios without modification or major retraining. A notable feature of the model is the ability to generate high-quality results without needing to specify the number of signal sources \emph{a priori} or the region of the coordinate space which is background-dominated.

However, the method relies heavily on access to large amounts of training data covering the entire range of scenarios the model will encounter, along with sufficient training time for the model. Unlike in the work of Ref.~\cite{Ullmo:2024odq}, no constraints or assumptions were needed to ensure training converged, nor were any issues with training encountered where weighting models provided any significant benefits. In conclusion, this method provides a powerful and effective approach for VHE gamma-ray signal-background separation using non-parametric reconstructions on a per pixel level and good error estimates based on model uncertainty. This model also provides a robust framework which can be used by future works to improve upon the results presented in this work and further expand its application to more complex astrophysics scenarios. In particular, more realistic data sets and regression of the SB value as a model output should be explored with particular interest.

\bibliographystyle{JHEP}
\bibliography{references}

@book{GonzalezWoods2018,
  author    = {Rafael C. Gonzalez and Richard E. Woods},
  title     = {Digital Image Processing},
  edition   = {4},
  publisher = {Pearson},
  year      = {2018}
}

@book{Goodfellow-et-al-2016,
    title={Deep Learning},
    author={Ian Goodfellow and Yoshua Bengio and Aaron Courville},
    publisher={MIT Press},
    note={\url{http://www.deeplearningbook.org}},
    year={2016}
}

@misc{HESSpublicrelease,
howpublished = {\url{https://www.mpi-hd.mpg.de/hfm/HESS/}}
}

@article{HESS:2022ygk,
    author = "Abdalla, H. and others",
    collaboration = "H.E.S.S.",
    title = "{Search for Dark Matter Annihilation Signals in the H.E.S.S. Inner Galaxy Survey}",
    eprint = "2207.10471",
    archivePrefix = "arXiv",
    primaryClass = "astro-ph.HE",
    doi = "10.1103/PhysRevLett.129.111101",
    journal = "Phys. Rev. Lett.",
    volume = "129",
    number = "11",
    pages = "111101",
    year = "2022"
}

@article{wang2022galaxy,
  title={Galaxy deblending using residual dense neural networks},
  author={Wang, Hong and Sreejith, Sreevarsha and Slosar, An{\v{z}}e and Lin, Yuewei and Yoo, Shinjae},
  journal={Physical Review D},
  volume={106},
  number={6},
  pages={063023},
  year={2022},
  publisher={APS}
}

@article{burke2019deblending,
  title={Deblending and classifying astronomical sources with Mask R-CNN deep learning},
  author={Burke, Colin J and Aleo, Patrick D and Chen, Yu-Ching and Liu, Xin and Peterson, John R and Sembroski, Glenn H and Lin, Joshua Yao-Yu},
  journal={Monthly Notices of the Royal Astronomical Society},
  volume={490},
  number={3},
  pages={3952--3965},
  year={2019},
  publisher={Oxford University Press}
}

@article{Schmidhuber:2014bpo,
    author = {Schmidhuber, J{\"u}rgen},
    title = "{Deep learning in neural networks: An overview}",
    doi = "10.1016/j.neunet.2014.09.003",
    journal = "Neural Networks",
    volume = "61",
    pages = "85--117",
    year = "2015"
}

@article{Kullback:1951zyt,
    author = "Kullback, S. and Leibler, R. A.",
    title = "{On Information and Sufficiency}",
    doi = "10.1214/aoms/1177729694",
    journal = "The Annals of Mathematical Statistics",
    volume = "22",
    number = "1",
    pages = "79--86",
    year = "1951"
}

@misc{kingma2017adammethodstochasticoptimization,
      title={Adam: A Method for Stochastic Optimization}, 
      author={Diederik P. Kingma and Jimmy Ba},
      year={2017},
      eprint={1412.6980},
      archivePrefix={arXiv},
      primaryClass={cs.LG},
      url={https://arxiv.org/abs/1412.6980}, 
}

@ARTICLE{726791,
  author={Lecun, Y. and Bottou, L. and Bengio, Y. and Haffner, P.},
  journal={Proceedings of the IEEE}, 
  title={Gradient-based learning applied to document recognition}, 
  year={1998},
  volume={86},
  number={11},
  pages={2278-2324},
  doi={10.1109/5.726791}}

@incollection{NIPS2012_4824,
  author = {Krizhevsky, Alex and Sutskever, Ilya and Hinton, Geoffrey E.},
  booktitle = {Advances in Neural Information Processing Systems 25},
  editor = {Pereira, F. and Burges, C. J. C. and Bottou, L. and Weinberger, K. Q.},
  pages = {1097--1105},
  publisher = {Curran Associates, Inc.},
  title = {ImageNet Classification with Deep Convolutional Neural Networks},
  url = {http://papers.nips.cc/paper/4824-imagenet-classification-with-deep-convolutional-neural-networks.pdf},
  year = 2012
}

@article{DBLP:journals/corr/abs-1906-02691,
  author       = {Diederik P. Kingma and
                  Max Welling},
  title        = {An Introduction to Variational Autoencoders},
  journal      = {CoRR},
  volume       = {abs/1906.02691},
  year         = {2019},
  url          = {http://arxiv.org/abs/1906.02691},
  eprinttype   = {arXiv},
  eprint       = {1906.02691},
  bibsource    = {dblp computer science bibliography, https://dblp.org}
}

@article{kingma2013autoencoding,
  title={Auto-encoding variational bayes},
  author={Kingma, Diederik P and Welling, Max},
  journal={arXiv preprint arXiv:1312.6114},
  year={2013}
}

@article{Rowell:2003jb,
    author = "Rowell, Gavin P.",
    title = "{A new template background estimate for source searching in TeV gamma-ray astronomy}",
    eprint = "astro-ph/0310025",
    archivePrefix = "arXiv",
    doi = "10.1051/0004-6361:20031194",
    journal = "Astron. Astrophys.",
    volume = "410",
    pages = "389",
    year = "2003"
}

@article{HESS:2018pbp,
    author = "Abdalla, H. and others",
    collaboration = "HESS",
    title = "{The H.E.S.S. Galactic plane survey}",
    eprint = "1804.02432",
    archivePrefix = "arXiv",
    primaryClass = "astro-ph.HE",
    doi = "10.1051/0004-6361/201732098",
    journal = "Astron. Astrophys.",
    volume = "612",
    pages = "A1",
    year = "2018"
}

@article{Berge:2006ae,
    author = "Berge, David and Funk, S. and Hinton, J.",
    title = "{Background Modelling in Very-High-Energy gamma-ray Astronomy}",
    eprint = "astro-ph/0610959",
    archivePrefix = "arXiv",
    reportNumber = "SLAC-PUB-12185",
    doi = "10.1051/0004-6361:20066674",
    journal = "Astron. Astrophys.",
    volume = "466",
    pages = "1219--1229",
    year = "2007"
}

@article{Ullmo:2024odq,
    author = "Ullmo, Marion and Moulin, Emmanuel",
    title = "{Nonparametric signal separation in very-high-energy gamma ray observations with probabilistic neural networks}",
    eprint = "2407.01329",
    archivePrefix = "arXiv",
    primaryClass = "astro-ph.HE",
    doi = "10.1088/1475-7516/2025/01/014",
    journal = "JCAP",
    volume = "01",
    pages = "014",
    year = "2025"
}

@misc{HESS,
howpublished = {\url{https://www.mpi-hd.mpg.de/HESS/}}
}

@misc{AdamSettings,
howpublished = {\url{https://docs.pytorch.org/docs/2.12/generated/torch.optim.Adam.html}}
}

@misc{CTAO,
howpublished = {\url{https://www.ctao.org/}}
}

\end{document}